\documentclass[prl,twocolumn,showpacs,amsmath,superscriptaddress,psfig,amssymb]{revtex4}

\usepackage{graphicx}
\usepackage{dcolumn}
\usepackage{bm}

\begin{document}


\title{S-wave/spin-triplet order in superconductors without inversion symmetry: Li$_2$Pd$_3$B and Li$_2$Pt$_3$B}

\author{H. Q. Yuan}
\email{yuan@mrl.uiuc.edu.}
\affiliation {Department of Physics,
University of Illinois at Urbana and Champaign, 1110 West Green
Street, Urbana, IL 61801}

\author{D. F. Agterberg}
\affiliation {Department of Physics, University of
Wisconsin-Milwaukee, Milwaukee, WI 53201}

\author{N. Hayashi}
\affiliation {Theoretische Physik, ETH-H\"{o}nggerberg, CH-8093
Z\"urich, Switzerland}

\author{P. Badica}
\affiliation {Institute for Materials Research, Tohoku
University, 2-1-1 Katahira, Aoba-ku, Sendai, 980-8577, Japan}
\affiliation {National Institute of Materials Physics, Bucharest,
P. O. Box MG-7, 077125, Romania}

\author{D. Vandervelde}
\affiliation {Department of Physics, University of Illinois at
Urbana and Champaign, 1110 West Green Street, Urbana, IL 61801}

\author{K. Togano}
\affiliation {National Institute for Materials Science, 1-2-1
Sengen, Tsukuba, 305-0047, Japan}

\author{M. Sigrist}
\affiliation {Theoretische Physik, ETH-H\"{o}nggerberg, CH-8093
Z\"urich, Switzerland}

\author{M. B. Salamon}
\affiliation {Department of Physics, University of Illinois at
Urbana and Champaign, 1110 West Green Street, Urbana, IL 61801}

\date{\today}

\begin{abstract}
\noindent We investigate the order parameter of noncentrosymmetric
superconductors Li$_2$Pd$_3$B and Li$_2$Pt$_3$B via the behavior
of the penetration depth $\lambda(T)$. The low-temperature
penetration depth shows BCS-like behavior in Li$_2$Pd$_3$B, while
in Li$_2$Pt$_3$B it follows a linear temperature dependence. We
propose that broken inversion symmetry and the accompanying
antisymmetric spin-orbit coupling, which admix spin-singlet and
spin-triplet pairing, are responsible for this behavior. The
triplet contribution is weak in Li$_2$Pd$_3$B, leading to a wholly
open but anisotropic gap. The significantly larger spin-orbit
coupling in Li$_2$Pt$_3$B allows the spin-triplet component to be
larger in Li$_2$Pt$_3$B, producing line nodes in the energy gap as
evidenced by the linear temperature dependence of $\lambda(T)$.
The experimental data are in quantitative agreement with theory.
\end{abstract}

 \maketitle




The crystal structure of most superconducting materials
investigated to date includes a center of inversion. The Pauli
principle and parity conservation then dictate that
superconducting pairing states with even parity are necessarily
spin-singlet, while those with odd parity must be spin-triplet
\cite{MUMM 2003}. In materials that lack inversion symmetry, the
tie between spatial symmetry and the Cooper-pair spin may be
broken \cite{Gorkov 2001, Frigeri 2004, Frigeri 2005, Edelstein
1989, Levitov 1985, Samokhin 2004}. The absence of inversion
symmetry along with parity-violating antisymmetric spin-orbit
coupling (ASOC) allows admixture of spin-singlet and spin-triplet
components. \ Unconventional behavior, including zeroes in the
superconducting gap function, is then possible, even if the pair
wavefunction exhibits the full spatial symmetry of\ the crystal.

In this Letter we report the dramatically different electrodynamic
behavior of two newly discovered noncentrosymmetric superconductors Li$_{2}$Pd$_{3}$B and Li$%
_{2}$Pt$_{3}$B \cite{Togano 2004, Badica 2005}. The penetration depth $%
\lambda (T)$ in the former material has the expected exponential
temperature dependence of a fully-gapped superconductor while the
latter exhibits a linear temperature dependence over the range
$0.05\leq T/T_{c}\leq 0.3.$ Inasmuch as the main difference
between these two compounds is the larger spin-orbit coupling
strength for Pt ((Z$_{Pt}$/Z$_{Pd}$)$^{2}$ = 2.9), we argue that
the unconventional behavior is evidence for admixed singlet and
triplet order as a consequence of ASOC. Indeed, we show
quantitative agreement between the experimental data of
$\lambda(T)$ and the theoretical calculations for mixed singlet
and triplet states based on ASOC.

Parity-broken superconductivity (SC) was previously discussed in
the context of surface superconductors \cite{Edelstein 1989} and
for dirty bulk materials \cite{Levitov 1985}. Recently, the
discovery of SC in the magnetic compounds CePt$_{3}$Si \cite{Bauer
2004}, UIr \cite{Akazawa 2004} and CeRhSi$_3$ \cite{Kimura
2005}(under pressure) has attracted extensive interest in studying
SC without inversion symmetry. Unfortunately, in these correlated
electron systems the nature of superconductivity is complicated by
its coexistence with magnetism, therefore severely restricting the
study of parity-broken SC.

Li$_{2}$Pd$_{3}$B and Li$_{2}$Pt$_{3}$B crystallize in a
perovskite-like cubic structure composed of distorted octahedral units of BPd$_{6}$ and BPt$%
_{6}$ \cite{Eibenstein 1997}. Unlike CePt$_3$Si, CeRhSi$_3$ and
UIr, these materials show no evidence of magnetic order or strong
correlated-electron effects  \cite{Togano 2004, Badica 2005,
Nishiyama 2005, Takeya 2005} that could lead to unconventional
superconducting behavior. Further, the increased spin-orbit
coupling in Pt leads to much larger band-splitting in
Li$_{2}$Pt$_{3}$B than in Li$_{2}$Pd$_{3}$B \cite{Lee 2005},
allowing us to study the dependence of superconductivity on the
ASOC strength. Therefore, we argue that Li$_{2}$Pd$_{3}$B and
Li$_{2}$Pt$_{3}$B provide a \textit{model system} in which to
study SC without inversion symmetry.

Polycrystalline samples of Li$_2$Pd$_3$B and Li$_2$Pt$_3$B were
prepared by arc melting \cite{Togano 2004, Badica 2005}. Powder
x-ray diffraction and metallography identify them as being single
phase. The sharp superconducting transitions with a width less
than 10\% of $T_c$ observed in either bulk
magnetization $M(T)$ (see, e.g., the inset of Fig.1), penetration depth $%
\lambda(T)$ or electrical resistivity $\rho(T)$ (not shown)
indicate good sample homogeneity. The mean free path \cite{note}, estimated from the rf resistivity $%
\rho$ at $T_c$, coherence length $\xi$ and specific heat
coefficient $\gamma$ \cite{Takeya 2005} ($\rho=20\mu\Omega$cm,
$\xi=9.5$nm, $\gamma=9$mJ/molK$^2$ for Li$_2$Pd$_3$B and $\rho=28 \mu\Omega$cm, $\xi=14.5$nm, $\gamma=7$mJ/molK$%
^2$ for Li$_2$Pt$_3$B), is 24 nm for Li$_2$Pd$_3$B and 42 nm for
Li$_2$Pt$_3$B, a few times larger than the corresponding
coherence length, indicating clean samples. Precise measurements of penetration depth $%
\Delta\lambda(T)$ were performed utilizing a tunnel-diode based,
self-inductive technique at 21 MHz down to 90 mK in a dilution
refrigerator. The change in penetration depth $\Delta\lambda(T)$
is proportional to the resonant frequency shift $\Delta f(T)$,
i.e., $\Delta \lambda(T)=G\Delta f(T) $, where the factor $G$ is
determined by sample and coil geometries \cite{Chia 2003}. Due to the uneven sample surface, the uncertainty of the $%
G $-factor can be up to 15\%. In this paper, $\Delta\lambda(T)$ is
extrapolated to zero at $T=0$, i.e., $\Delta\lambda(T)=\lambda(T)-\lambda_0$%
. The values of zero temperature penetration depth $\lambda_0$ ($%
\lambda_0=190$ nm for Li$_2$Pd$_3$B and $\lambda_0=364$ nm for Li$_2$Pt$_3$%
B) are taken from Ref.~\cite{Badica 2005}, determined from the
magnetic critical field measurements. The difference of
$\lambda_0$ in the two compounds might result from their distinct
Fermi surfaces due in part to the spin-orbit coupling \cite{Lee
2005}. The magnetization $M(T, H)$ was measured using a commercial
SQUID magnetometer (MPMS, Quantum Design).

\begin{figure}[tbp]
\centering \includegraphics[width=0.76\columnwidth]{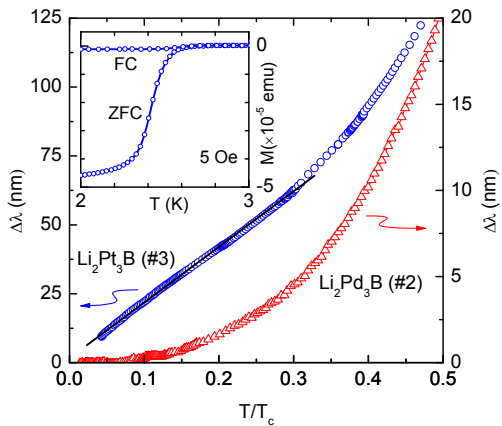}
\caption{(Color online). Temperature dependence of the penetration depth $%
\triangle\protect\lambda(T)$ for Li$_2$Pd$_3$B (\#2) and
Li$_2$Pt$_3$B (\#3), showing distinct low-temperature behavior
\protect\cite{Yuan 2005}. The inset shows the magnetization $M(T)$
for Li$_2$Pt$_3$B (\#3) measured in zero-field-cooling (ZFC) and
field-cooling (FC) in a magnetic field of 5 Oe. The values of
$T_c$ ($T_c=6.7$ K for Li$_2$Pd$_3$B (\#2) and $T_c=2.43$ K for
Li$_2$Pt$_3$B (\#3)) were determined from the mid-points of
magnetization drop at $T_c$.}
\end{figure}

In Fig.\ 1 the penetration depth change $\Delta \lambda (T)$ is shown for Li$%
_{2}$Pd$_{3}$B and Li$_{2}$Pt$_{3}$B, respectively. The nearly $T$%
-independence of $\lambda (T)$ at low temperatures for the
Pd-compound is characteristic of fully gapped behavior, consistent
with NMR experiments \cite{Nishiyama 2005} and specific heat
measurements \cite{Takeya 2005}. However, the penetration depth
$\lambda (T)$ of Li$_{2}$Pt$_{3}$B follows a linear temperature
dependence \cite{Yuan 2005}. Such a $T$-linear behavior of
$\lambda(T)$ can be theoretically interpreted by (a) phase
fluctuations among Josephson-coupled grains \cite{Ebner 1983} and
(b) line nodes in the superconducting energy gap. The former one
can be ruled out in this context. The importance of phase
fluctuations depends inversely on grain size, which is large ($>
$100 $\mu $m) in both the Pt- and Pd-samples \cite{Togano 2004}.\
If phase fluctuations dominate, the Pd sample, with comparable
normal-state resistivity and grain size, should also show a
strong linear temperature dependence. Further, the transition
temperature $T_c$ is strongly dependent on the normal state
resistivity in the phase-fluctuation regime. We find that the
$T_c$ varies by less than 10\% among samples that have
normal-state resistivities that differ by a factor of
three or more. Finally we have reanalyzed the specific heat of Li$_{2}$Pt$%
_{3}$B reported in Ref.\cite{Takeya 2005} and find that
$C_{el}/T\sim T$ \ is a much better representation of those data
at low temperature than is an exponential dependence, further
supporting the existence of line nodes.

Before describing our model, we explore possible non s-wave states
that might exhibit line nodes in Li$_2$Pt$_3$B. Weak-coupling theory of SC, justified by the low $%
T_{c}$, permits only the following three:\\
(i) $\Delta_{+}(\mathbf{k})\simeq\Delta_{-}(\mathbf{k})=(k_x^2-k_y^2)(k_y^2-k_z^2)(k_z^2-k_x^2)$,\\
(ii)$\Delta_{+}(\mathbf{k},z)\simeq\Delta_{-}(\mathbf{k},z)=e^{iqz}k_z[k_y(k_y^2-k_z^2)+ik_x(k_z^2-k_x^2)]$,\\
(iii) $\Delta_{+}(\mathbf{k},z)\simeq\Delta_{-}(\mathbf{k},z)=e^{iqz}k_z(k_x+ik_y)$.\\
In the latter two cases, broken parity and time reversal
symmetries combine to destabilize the spatially uniform state,
giving rise to the spatial dependence in the gap functions. The
former two states are unlikely in any theory that is based on
local interactions (like the single band Hubbard model). Since the
above three states are not s-wave pairing states, and (as argued
below) Li$_{2}$Pd$_{3}$B appears to be s-wave, a phase transition
in the pairing state of Li$_{2}$(Pd$_{1-x}$Pt$_{x}$)$_{3}$B with
varying $x$ would have to occur for any of these states to exist
in Li$_{2}$Pt$_{3}$B. Furthermore, these states should be extremely sensitive to impurities and $%
T_{c}$ should be strongly suppressed when $x$ is varied away from
1. These are in contrast with the experimentally observed smooth
evolution of $T_{c}$ with $x$ \cite{Badica 2005}. Given these
arguments against unconventional superconductivity, we attribute
the dramatic difference between these two compounds to the
variation of ASOC.

When parity symmetry is violated, the ASOC that breaks the spin
degeneracy of each band takes the form $\alpha\mathbf{g(k)\cdot S(k)}/\hbar$, where $%
\alpha$ denotes the spin-orbit coupling strength, $\mathbf{S(k)}$
is the spin of an electron with momentum $\hbar \mathbf{k}$, and
$\mathbf{g(k)}$ is a dimensionless vector ($\mathbf{g(-k)=-g(k)}$
to preserve time reversal symmetry). This ASOC leads to an energy
splitting of the originally degenerate spin states and results in
spin-eigenstates that are polarized parallel or anti-parallel to
$\mathbf{g(k)}$. The ASOC plays a crucial role in the
determination of the superconducting state. The key point is that
if a spin-triplet contribution to the superconducting gap function
is to emerge, its characteristic d-vector $\mathbf{d(k)}$ must be parallel to $%
\mathbf{g(k)}$ (provided that the ASOC is sufficiently large)
\cite{Frigeri 2004, Frigeri 2005}. This leads to two gap functions $\Delta_{\pm}(\mathbf{k}%
)=\psi\pm t \mid\mathbf{g(k)}\mid$, where each gap is defined on
one of the two bands formed by the degeneracy lifting of the ASOC;
$\psi$ and $t$ are the singlet and triplet order parameters
respectively. For a range of values of $\nu=\psi/t$,
$\Delta_{-}(\mathbf{k})$ can change sign and nodes may exist in
the superconducting gap.

Recent band structure calculations for these compounds \cite{ Lee
2005} provide information about $\mid\mathbf{g(k)}\mid$. These
results indicate that $\alpha$ is a large energy scale relative to the bandwidth and that $%
\mid\mathbf{g(k)}\mid$ is highly anisotropic. To capture these
results in a model we take:\newline
$\mathbf{g(k)}=a_1\mathbf{k}-a_2[\widehat{\mathbf{x}}k_x(k_y2+k_z2)+%
\widehat{\mathbf{y}}k_y(k_z2+k_x2)+\widehat{\mathbf{z}}k_z(k_x2+k_y2)]$,%
\newline
with $a_2/a_1=3/2$, $\mathbf{k}$, a unit vector, and the spherical
average of $\mid\mathbf{g(k)}\mid^2$ equal to unity. This form of $\mid\mathbf{g(k)}%
\mid$ is the simplest that is consistent with cubic symmetry and
allows for anisotropy on a model, spherical Fermi surface.

\begin{figure}[tbp]
\centering \includegraphics[width=0.75\columnwidth]{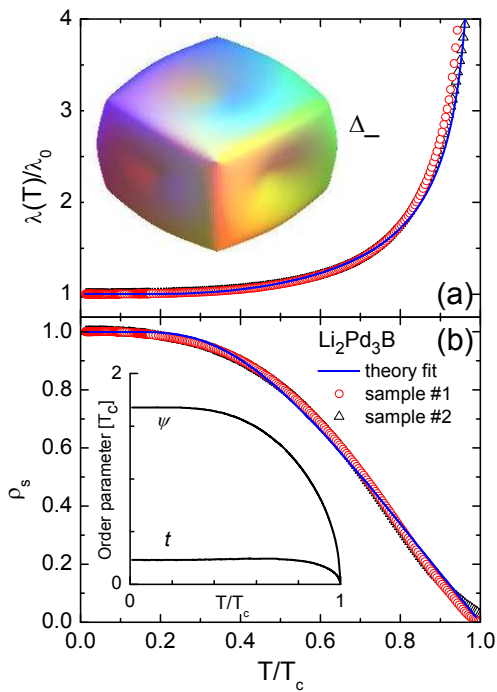}
\caption{(Color online). The temperature dependence of (a) the
normalized penetration depth
$\protect\lambda(T)/\protect\lambda_0$ and (b) the corresponding
superfluid density $\protect\rho_s(T)$ for Li$_2$Pd$_3$B, in which
$T_c=7$ K, $G=0.42$ nm/Hz for sample \#1 and $T_c=6.7$ K, $G=0.63$
nm/Hz for sample \#2. The symbols, as described in the figure,
represent the experimental data and the solid line is a theoretical fit with parameters $%
\protect\delta=0.1$ and $\protect\nu=4$. The insets in the upper
panel and the lower panel show a 3-dimensional (3D) polar plot of the gap function $%
\Delta_{-}(\mathbf{k})$, and the temperature dependence of the
order parameter components $\protect\psi$ (spin singlet) and $t$
(spin triplet), respectively. }
\end{figure}

\begin{figure}[tbp]
\centering \includegraphics[width=0.75\columnwidth]{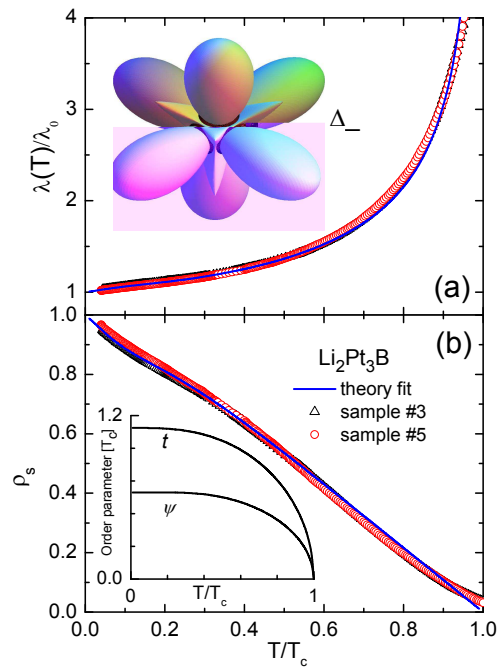}
\caption{(Color online). The temperature dependence of (a) the
normalized penetration depth
$\protect\lambda(T)/\protect\lambda_0$ and (b) the corresponding
superfluid density $\protect\rho_s(T)$ for Li$_2$Pt$_3$B, in which
$T_c=2.43$ K, $G=1$ nm/Hz for sample \#3 and $T_c=2.3$ K, $G=0.41$
nm/Hz for sample \#5. The fitting parameters are $\protect\delta=0.3$ and $%
\protect\nu=0.6$. In Li$_2$Pt$_3$B, the spin-triplet component $t$
is the dominant order parameter (the inset of Fig.\ 3(b)). In
order to clearly show the line nodes, a small constant is added to the gap function $\Delta_{-}(%
\mathbf{k})$ (the inset of Fig.\ 3(a)). Six circle-like line nodes
can be seen along the large lobes as marked by the dark lines. $\Delta_{-}(\mathbf{k%
})$ changes sign from the large lobes (+) to the small lobes (-)
in the 3D polar plot.}
\end{figure}

We compute the penetration depth (the superfluid density) on the
basis of the formula described in \cite{Hayashi 2005}. These fits
provide estimates for $\nu $ (defined at $T\rightarrow T_{c}$) and
$\delta $, the ratio of the relative density of states between the
spin-orbit-split bands. The resulting fits are shown in Figs.\
2(a) and 3(a), respectively. Li$_{2}$Pd$_{3}$B is nearly a pure
spin-singlet state, with a large value of $\nu \simeq 4$. We note
that the preliminary fit of two-band model in Li$_{2}$Pd$_{3}$B
with a fraction of 4\% from the small energy gap \cite{Yuan 2005}
treated data only for $T<0.3T_{c}$, while the present analysis
covers the whole temperature range. As argued above,
Li$_{2}$Pt$_{3}$B clearly evidences line nodes, meaning that
$\Delta _{-}(\mathbf{k})$ changes sign for a range of wavevectors.
The best fit for Li$_{2}$Pt$_{3}$B has $\nu =0.6$ and $\delta
=0.3$, indicating that the spin-triplet component is dominant. We expect $%
\delta $ to be proportional to the strength $\alpha $ of the ASOC,
which in turn varies as the square of the atomic number, as above.
The obtained value of $\delta (Pt)/\delta (Pd)=0.3/0.1=3$ is
consistent with the expectations. In the insets to Figs.\ 2(a) and 3(a) we show polar plots of $\Delta _{-}(%
\mathbf{k})$ for the two compounds; for Li$_{2}$Pt$_{3}$B, the
existence of line nodes appears in the form of circular bands. For
Li$_{2}$Pd$_{3}$B, both $\Delta _{+}(\mathbf{k})$ and $\Delta
_{-}(\mathbf{k})$ are non-zero, but anisotropic, over the entire
Fermi surface. It is noted that the gap functions $\Delta
_{+}(\mathbf{k})$ and $\Delta _{-}(\mathbf{k})$ possess cubic
symmetry (see the insets to Figs.\ 2(a) and 3(a)) and the pairing
states break only gauge invariance symmetry, i.e., an s-wave
orbital symmetry for both Li$_{2}$Pd$_{3}$B and Li$_{2}$Pt$_{3}$B.
However, $\Delta _{-}(\mathbf{k})$ exhibits a sign change in
Li$_{2}$Pt$_{3}$B indicated by dark circles in Fig. 3(a). Figures
2(b) and 3(b) present the superfluid density $\rho _{s}(T)$
obtained from the penetration depth ($\rho _{s}(T)=\lambda
_{0}^{2}/\lambda ^{2}(T)$), along with calculated curves. The
agreement is satisfactory. One notes that the weak tail in the
experimental $\rho _{s}(T)$ as $T\rightarrow T_{c}$ is mainly due
to the influence of rf skin depth upon approaching $T_{c}$. The
insets to Figs.\ 2(b) and 3(b) show the calculated temperature
dependences of the order parameters $\psi $ and $t$. Obviously,
the spin-singlet component is dominant in the order parameter of
Li$_{2}$Pd$_{3}$B, but it is not the case in Li$_{2}$Pt$_{3}$B.
For the latter compound, the spin-triplet component $t$ is
sufficiently large to give rise to the existence of line nodes in
the superconducting energy gap. The existence of a spin-triplet
state may be stabilized by the \textquotedblleft inter-parity"
coupling (termed $e_{m}$ in Ref.\cite{Frigeri 2005}) between
singlet and triplet channels as allowed by broken inversion
symmetry. This interaction can arise from el.-ph. (and el.-el.)
coupling and may dominate in Li$_{2}$Pt$_{3}$B because of the
large ASOC \cite{Lee 2005}. We note that while our model
(spherical Fermi surface and isotropic spin-singlet gap) predicts
that spin-triplet component is larger than the spin-singlet
component, this needs not be the case in reality. In particular,
if the spin-singlet gap is anisotropic, then the Fermi surface
average of the magnitude of the spin-triplet component required to
produce line nodes can be significantly decreased.

In addition to the profound effect on the pairing state in
Li$_2$Pt$_3$B, broken parity symmetry has other non-trivial
consequences. For example, the cubic symmetry allows for a novel
contribution to the Ginzburg Landau (GL)
free energy density of the form $\varepsilon\mathbf{B \cdot j_{so}}$, where $%
\mathbf{j_{so}}$ is the supercurrent as defined in the usual GL theory and $%
\varepsilon$ is a constant. As a consequence, the condensate
wavefunction will not be spatially uniform along the direction of
the applied magnetic field as it usually is. Near the upper
critical field, it will develop a
finite center of mass momentum that is parallel to the applied field \cite%
{Kaur 2005}. This helical structure of the order parameter is
similar to that of a Fulde-Ferrell-Larkin-Ovchinnikov (FFLO)
superconductor \cite{Fulde 1964, Larkin 1965}. However, in
contrast to the FFLO phase, a non-zero center of mass momentum
exists at all temperatures. In the vortex state, this coupling
term causes the magnetization to develop a transverse component
that is parallel to the supercurrent. This may be observable
through small angle neutron scattering experiments.

In summary, our observations have demonstrated that
superconductors lacking inversion symmetry exhibit qualitatively
distinct properties from those with an inversion center. The
existence of unconventional properties (e.g., line nodes) for an
s-wave type superconductor, found in Li$_2$Pt$_3$B, provides an
alternative way to study unconventional SC, especially that
arising from phonon pairing mechanism. Indeed, the absence of
parity symmetry coupled with strong spin-orbit coupling, which
results in an admixture of spin-singlet and spin-triplet pairing,
requires a complete reconceptualization of Cooper pairs and the
nature of the superconducting state.

We thank P. Frigeri, R. Kaur, I. Milat and H. Takeya for useful
discussions. This work is supported by the National Science
Foundation under awards No. NSF-EIA0121568 and NSF-DMR0318665, the
Department of Energy under award No. DEFG02-91ER45439, the Swiss
National Science Foundation and Petroleum Research Funds. We also
acknowledge supports from ICAM (HQY), the 2003-PFRA program of
Japan Society for the Promotion of Science (NH) and the CEEX
program of Romanian Ministry of Education and Research (PB).

\end{document}